%% file: arxiv-v1.tex
\documentclass[aps,twocolumn,nofootinbib,groupedaddress]{revtex4-1}

\usepackage[english]{babel}
\usepackage{graphicx}
\usepackage{amsmath,amssymb,amsbsy,amstext, amsthm, simplewick, amsfonts}
\usepackage{dcolumn}

\usepackage{subfiles}

\usepackage{hyperref}
\hypersetup{
    colorlinks=true,
    linkcolor=purple,
    filecolor=purple,      
    urlcolor=purple,
    citecolor=purple
    }
    
\usepackage{comment}
\usepackage{bm}
\usepackage{xcolor}
\usepackage{braket}
\usepackage{tensor}
\usepackage{ulem}

\newcommand{\dd}{\mathrm{d}}

\def\Mp{M_{\rm pl}}
\def\F{\mathcal{F}}
\def\S{\mathcal{S}}

\begin{document}


\title{The effective field theory of multi-field inflationary fluctuations
}

\author{Lucas Pinol}
\email{lucas.pinol@phys.ens.fr}
\affiliation{Laboratoire de Physique de l’École Normale Supérieure, ENS, CNRS, Université PSL,\\ Sorbonne Université, Université Paris Cité, F-75005, Paris, France}

\begin{abstract}
We build an effective field theory of multi-field inflationary fluctuations based on the adiabatic perturbation and on any number of matter fluctuations in the non-adiabatic sector, without imposing extra symmetries on the latter. Focusing on terms with at most two derivatives in fields' fluctuations, we argue that taking the decoupling limit---in which gravitational interactions are neglected---is justified in a quasi de Sitter spacetime with slow-varying Hubble scale. With these working hypotheses, we find simple forms of multi-field mixings (quadratic order) and interactions (cubic order). We explain how to break degeneracies amongst various terms, and we compare the predictions of the effective field theory to those of non-linear sigma models of inflation and more general multi-field Lagrangian in the traditional model approach. We stress that several multi-field cubic interactions are dictated by non-linearly realised spacetime symmetries and are therefore given in terms of parameters already present in the quadratic action. We propose various directions to systematically explore the phenomenology generic to multi-field inflation and beyond the lamppost of known models.
\end{abstract}

\maketitle

\paragraph*{\bf Introduction.} 

Effective field theories (EFT) are powerful tools to systematically build physical theories beyond the lamppost of known models.
In the inflationary context, the first stones were laid in~\cite{Creminelli:2006xe,Cheung:2007st} with the proposition of an EFT describing the adiabatic degree of freedom, ever present in cosmology.
This elegant formalism evades possible criticisms to the model approach to inflation, which in its paradigmatic expression postulates the existence of an ad hoc fundamental scalar field.
Delegating the details of the background spacetime dynamics to the specification of time-dependent Wilson-like coefficients to be determined by experiments, the EFT treatment focuses on symmetries to systematically build a Lagrangian to any order in perturbations and following consistent approximation schemes.

Quite surprisingly, an EFT for inflation in the general context of multiple matter fluctuations is not available in the literature so far.
The first notable work in this direction was~\cite{Senatore:2010wk}, but many of the interesting multi-field effects are absent because an additional shift symmetry was imposed on the matter sector, as emphasized in~\cite{Shiu:2011qw}.
This hole was partially filled by~\cite{Noumi:2012vr}, proposing an interesting EFT description of the then recently unveiled Quasi-Single-Field scenario~\cite{Chen:2009zp}, in which a single additional matter fluctuation features a mass term and couples non-derivatively with the adiabatic sector.
This paved the way to an effective description of two-field inflation, implicitly encompassed in many of the subsequent works dedicated to the study of the oscillatory signal discovered in the squeezed limit of the primordial bispectrum in this context~\cite{Noumi:2012vr}, shortly after dubbed as the cosmological collider signal~\cite{Arkani-Hamed:2015bza} for the frequency is mostly set by the mass of the matter fluctuation (see~\cite{Pinol:2023oux} for an in-depth discussion of how the quadratic mixing affects this frequency).

In the recent years, and despite impressive progresses in computing correlators in multi-field inflation as illustrated by the surge of bootstrap techniques in cosmology~\cite{Arkani-Hamed:2018kmz}, only very recently was the cosmological collider signal shown, for any number of massive fields, to lead to a striking new phenomenon dubbed as inflationary flavour oscillations~\cite{Pinol:2021aun}.
Because no general EFT for inflationary fluctuations was available at the time, the latter reference focuses on known quadratic and cubic Lagrangians for fluctuations~\cite{Pinol:2020kvw} 
in the large class of non-linear sigma models (NLSM), that may be encountered in low-energy predictions from tentative fundamental physical theories like string theory~\cite{Baumann:2014nda}.
However, more general multi-field models
may also be predicted by motivated UV completions, as exemplified by the multi-field versions of the DBI scenario~\cite{Kehagias:1999vr,Steer:2002ng}.
It would be desirable to dispose of a model-independent description of multi-field interactions in order to derive generic predictions of multi-field inflation beyond the lamppost of known models.

In this work, we precisely build an EFT of inflation that allows for any number of fluctuations' species and does not assume additional symmetries in this matter sector.
Our explicit construction is truncated at two-derivatives order but it allows for straight generalisations to higher orders. We propose some EFT building blocks made of matter fluctuations and their derivatives in the unitary gauge, to be added and combined with usual operators for the adiabatic degree of freedom.
We prove that, even in this generic multi-field context, taking the decoupling limit is justified in a consistent slow-roll expansion.
Prompted by the knowledge of the consequences of non-linearly realised spacetime symmetries in the adiabatic sector, we identify a set of multi-field cubic interactions whose strengths are fully fixed by parameters in the quadratic action.
Our resulting EFT encompasses and generalises those of Refs.~\cite{Senatore:2010wk,Noumi:2012vr} and, when applicable, matches expectations from the known Lagrangians for fluctuations in non-linear sigma models~\cite{Pinol:2020kvw}.

\vskip 4pt
\paragraph*{\bf The traditional model approach.} 

The traditional model approach to inflation consists in providing a Lagrangian for the fundamental scalar fields $\phi^I(t,\vec{x})$.
Those are responsible both for the classical quasi-exponential expansion of the background spacetime, and for providing---via their quantum fluctuations---the seeds for the formation of the large scale structures in our universe.
The most popular multi-field models of inflation arguably lie in the large class of NLSM models, with the following action,
$$S_\mathrm{NLSM} = \int \sqrt{-g}\dd^4 x \left[-\frac{1}{2}G_{IJ}(\vec{\phi}) g^{\mu\nu} \partial_\mu\phi^I \partial_\nu \phi^J - V(\vec{\phi}) \right],$$
and minimally coupled to gravity as described by the Einstein-Hilbert action.
These models encompass both kinetic interactions via the field-space metric $G_{IJ}$, and potential interactions in $V$.
Their observational predictions have been thoroughly investigated in the past, with unique phenomena ranging, e.g., from suppressed tensor-to-scalar ratio in multi-field $\alpha$-attractors~\cite{Achucarro:2017ing} to (respectively transient) instabilities due to a negative contribution to the mass of entropic fluctuations in hyperbolic field spaces~\cite{Renaux-Petel:2015mga} (respectively due to a strongly non-geodesic motion of the background trajectory~\cite{Fumagalli:2019noh}), and many other ones.

Most of the aforementioned scenarios specify to the two-field case, for the physics is already rich enough due to the quadratic mixing between linear fluctuations.
This becomes even more evident in terms of the adiabatic-entropic decomposition~\cite{Gordon:2000hv}, in which one learns that the curvature fluctuation is sourced by the entropic one whenever the background trajectory does not follow a field-space geodesic.
Quadratic and cubic Lagrangians for fluctuations in the adiabatic-entropic basis for NLSM were first shown in~\cite{Garcia-Saenz:2019njm} in the two-field case, and quickly prompted the generalisation to any number of fluctuations in~\cite{Pinol:2020kvw}.
The extension is not just technical but also conceptual, as genuinely many-field effects may affect cosmological observables, as well illustrated by the discovery of the inflationary flavour and mass bases and related flavour oscillations~\cite{Pinol:2021aun}.
The quadratic action for the adiabatic fluctuation $\zeta$ and the non-adiabatic ones $\F^\alpha$ with $\alpha \in \{1,\ldots,N_\mathrm{field}-1\}$ reads $$S^{(2)}_\mathrm{NLSM} = \int \sqrt{-g} \,\dd^4 x\, \mathcal{L}^{(2)}_\mathrm{NLSM}\,,$$ with~\cite{Pinol:2020kvw}
\begin{align}
\label{eq: L2 model}
\mathcal{L}^{(2)}_\mathrm{NLSM}
 = \, & - \frac{1}{2} \frac{f_\pi^4}{H^2} g^{\mu\nu}  \partial_\mu \zeta \partial_\nu \zeta  - \frac{1}{2}g^{\mu\nu}   \partial_\mu \F^\alpha\partial_\nu \F ^\alpha \,,  \nonumber \\
  & -\frac{1}{2}m_{\alpha\beta}^2 \mathcal{F}^\alpha \mathcal{F}^\beta - \Omega_{\alpha \beta} g^{\mu 0}\partial_\mu \F^\alpha \F^\beta \,,  \nonumber 
 \\   & - 2 \frac{f_\pi^2}{H} \omega_1 \delta_{1\alpha} g^{\mu0} \partial_\mu \zeta  \F^\alpha\,, 
\end{align}
where $f_\pi^2 = \sqrt{2 \epsilon c_s} H \Mp$ represents the normalisation of the curvature fluctuation (with $c_s=1$ so far).
The typical size of $\zeta$ is indeed  $\Delta_\zeta \equiv A_s^{1/2} = H^2/(2\pi f_\pi^2)$, with $A_s = 2.1 \times 10^{-9}$~\cite{Planck:2018jri}.
In addition to kinetic terms with unit speeds of sound, $\mathcal{L}^{(2)}_\mathrm{NLSM}$ is composed of purely non-adiabatic mixing via the symmetric mass matrix $m^2_{\alpha\beta}$ and the anti-symmetric kinetic mixing $\Omega_{\alpha\beta}$, as well as an adiabatic-entropic mixing via $\omega_1$ and the single fluctuation $\F^1$ playing the role of a portal field~\cite{Pinol:2021aun}.
Notice that Lorentz covariance is explicitly broken at the level of fluctuations, as expected in a cosmological context.
In this traditional model approach, the couplings have a clear interpretation.
The geodesic deviation of the background trajectory is encoded in $\omega_1$ and the higher-order ``curvatures'' in $\Omega_{\alpha \beta}$.
Moreover, the mass matrix $m^2_{\alpha\beta}$ is expressed in terms of properties of the NLSM Lagrangian, including covariant derivatives of the potential and information about the geometry of the field space via its Riemann tensor.
The most salient features of the NLSM cubic interactions found in~\cite{Pinol:2020kvw} are reviewed in Appendix A.

Before moving to the EFT construction, it is worth noting the existence of multi-field models of inflation beyond NLSM.
A larger class features Lagrangian densities $P(X^{IJ},\vec{\phi})$ that are generic functions of both kinetic terms $X^{IJ} = -g^{\mu\nu} \partial_\mu \phi^I \partial_\nu\phi^J/2$ and of fields $\phi^I$, chief amongst which is the multi-field DBI scenario~\cite{Kehagias:1999vr,Steer:2002ng}.
Although quadratic and cubic Lagrangians in terms of adiabatic-entropic fluctuations for any number of fields in this larger class of models has not been elucidated yet, it is instructive to focus on the particular case of two-field DBI.
In addition to a non-unit speed of sound for $\zeta$ (as also in general single-field $P(X,\phi)$ scenarios~\cite{Garriga:1999vw}) and $\F^1$~\cite{Langlois:2008qf}, cubic interactions appear there with more derivatives than those found in NLSM~\cite{Langlois:2008wt}.

\vskip 4pt
\paragraph*{\bf EFT building blocks.} 
An EFT approach to inflationary fluctuations was first proposed in~\cite{Creminelli:2006xe,Cheung:2007st}.
Within this celebrated framework, the ever presence of the adiabatic mode in cosmology becomes evident as the existence of a pseudo-Nambu-Goldstone (NG) boson associated to the time diffeomorphism invariance broken by the cosmological background.
Since this is already well-established (see, e.g.,~\cite{Piazza:2013coa} for a review), we simply quote the general Lagrangian density for the adiabatic sector in the unitary gauge, in which the NG boson is eaten by the spacetime metric:
\begin{align}
\label{eq: Lad}
     \mathcal{L}_\mathrm{ad} = & \, \frac{\Mp^2}{2} R + \Mp^2 \dot{H} g^{ 00} - \Mp^2(3H^2 +\dot{H})
     \\\, & +  \sum_{n=2}^\infty\frac{M_{n}^4}{n!}\left(\delta g^{00}\right)^{n} +  \ldots\,,\nonumber
\end{align}
where an over-dot represents a derivative with respect to cosmic time $t$ and $\delta g^{00} = g^{00}+1$.
The first line is called the universal part of the EFT as it is uniquely determined by enforcing a flat FLRW background verifying Friedmann equations without spatial curvature.
The second line represents a tower of operators with time-dependent Wilson-like coefficients $M_n(t)$, to be determined by experiments or by matching with a specific UV-completion\footnote{
Two important examples are i) $P(X,\phi)$ single-field models, where $M_n$ depend on $P$ and its derivatives~\cite{Burrage:2011hd}; ii) NLSM after integrating out entropic fluctuations when they are all much heavier than the Hubble scale, where $M_n$ depend on diverse multi-field properties like masses, couplings and geometry of the field space~\cite{Pinol:2020kvw}.}. 
The dots represent terms at higher order in derivatives of the metric, as conveniently described by the extrinsic curvature perturbation $\delta K_{\mu\nu}=K_{\mu\nu}- H h_{\mu\nu}$ where $h_{\mu\nu}$ is the induced metric on the spatial hypersurfaces at constant $t$, as well as the corresponding intrinsic curvature $R^{(3)}_{ij}$.
These interesting terms, which always result in second-order equations of motion in time but possibly higher order in space, may have important effects and are described in Appendix B, but the main body of this manuscript will overlook them.

Although definitely useful to define blocks for building a general Lagrangian density, the variables appearing above can hardly be used to make concrete calculations.
We therefore introduce explicitly the NG boson $\pi(t,\vec{x})$ that, from the expectation that it should restore full diffeomorphism invariance, we know must transform as $\pi \rightarrow  \pi - \xi(t,\vec{x})$ under a time diffeomorphism $t \rightarrow t +\xi(t,\vec{x}) $, such that $t+\pi$ is invariant.
By analogy with particle physics, this is known as the Stückelberg trick.
To be specific, and this will be important when exploring the mixing with gravity and the decoupling limit, we define $\pi$ in the spatially flat gauge where $h_{ij} \rightarrow a^2\delta_{ij}$.
Under Stückelberg, our fundamental building block transforms as
\begin{align}
\label{eq: stuckelberg}
    t \rightarrow t+\pi \,\, \implies \,\, \delta g^{ 00} & \rightarrow \delta g^{ 00}_\mathrm{flat} - 2 \dot{\pi} - \dot{\pi}^2 + \frac{(\partial_i \pi)^2}{a^2} \,.
\end{align}
All time-dependent functions are shifted at the time $t+\pi$.
Although we assume that the Hubble scale and its derivatives are slowly varying, we discuss after Eq.~\eqref{eq: before time dependence discussion} important effects of time-dependent couplings from the viewpoint of non-linearly realised symmetries.
Note that four-dimensional scalars like $R$ and $\dd^4 x\sqrt{-g}$ transform covariantly under time diffeomorphisms and do not generate new terms.

We now turn to the non-adiabatic sector and suppose the existence of $N_\mathrm{field}-1$ matter fluctuations, denoted as $\mathcal{S}^\alpha$, and that covariantly transform as scalars under diffeomorphisms.
We come back to the unitary gauge to construct EFT building blocks respecting the subset of unbroken FLRW symmetries, and consider a derivative expansion up to second order only,
\begin{align}
\label{eq: Lnad}
    \mathcal{L}_\mathrm{nad,0} &= \sum_{n,k}\frac{b_{\alpha_1\ldots\alpha_k}^{(n,k)}}{n!k!}\left(\delta g^{00}\right)^n 
    \mathcal{S}^{\alpha_1}\ldots\mathcal{S}^{\alpha_k} \,,\\
    \mathcal{L}_\mathrm{nad,1} &= \sum_{n,k}\frac{c_{\alpha\alpha_1\ldots\alpha_k}^{(n,k)}}{n!k!}\left(\delta g^{00}\right)^n g^{0\mu}\partial_{\mu}\S^{\alpha} \S^{\alpha_1}\ldots \S^{\alpha_k} \nonumber \,, \\
    \mathcal{L}_\mathrm{nad,2} &= \sum_{n,k}\left(\delta g^{00}\right)^n \left(\bar{d}_{\alpha\beta\alpha_1\ldots\alpha_k}^{(n,k)}g^{0\mu}g^{0\nu}-d_{\alpha\beta\alpha_1\ldots\alpha_k}^{(n,k)}g^{\mu\nu} \right) \nonumber \\ 
    & \quad \,\,\times\frac{1}{2 }\frac{1}{n!k!}\partial_{\mu}\S^{\alpha} 
    \partial_\nu \S^{\beta}
    \S^{\alpha_1}
    \ldots \S^{\alpha_k} \nonumber \,,
\end{align}
where all $(b\,, c\,, d\,, \bar{d})$ tensors are time dependent and of mass dimensions $(4-k\,,2-k\,,-k\,,-k)$.
Additionally, $b$ is fully symmetric, $c$ is symmetric under any permutation of its last $k$ indices, and $d$ and $\bar{d}$ are symmetric with respect to the two first indices on one hand, and the $k$ last ones on the other hand.
The sums are on $(n,k) \in \mathbb{N}^2$, but we fix $b^{(n,0)}=0$ to avoid degeneracies with the coefficients $M_n$ in~\eqref{eq: Lad}.
Additional towers of operators at this derivative order can be constructed using the 3d curvatures and are discussed in Appendix B.

\vskip 4pt
\paragraph*{\bf Breaking degeneracies.}

Using the Stückelberg trick, we restore full diffeomorphism invariance, we specify to the spatially flat gauge and we regroup terms by their power in fields' fluctuations.
We use
\begin{align}
    g^{ 0 i}  \rightarrow g^{ 0 i}_\mathrm{flat} +\frac{\delta^{ij}}{a^2} \partial_j  \pi, \,
    \partial_0  \rightarrow \frac{1}{1+\dot{\pi}} \partial_0,\,
    \partial_i \rightarrow \partial_i -  \frac{\partial_i \pi}{1+\dot{\pi}} \partial_0, \nonumber
\end{align}
as well as $\S^\alpha \rightarrow \S^\alpha$.
By requiring tadpole cancellation, i.e. vanishing of $\mathcal{L}^{(1)}_\mathrm{nad}$, we find that $b_\alpha^{(0,1)} = c_\alpha^{(0,0)} =0$, which will have important consequences when discussing non-linearly realised symmetries.
The quadratic Lagrangian density is found only from the following terms in the unitary gauge,
\begin{align}
    &\frac{b_{\alpha\beta}^{(0,2)}}{2}\S^\alpha\S^\beta + b_\alpha^{(1,1)} \delta g^{00} \S^\alpha  + c_{\alpha\beta}^{(0,1)} g^{0\mu} \partial_\mu \S^\alpha\S^\beta + c_\alpha^{(1,0)} \times \nonumber\\ &  \delta g^{00} g^{0\mu} \partial_\mu \S^\alpha + \frac{1}{2} \left(\bar{d}_{\alpha\beta}^{(0,0)}g^{0\mu}g^{0\nu}-d_{\alpha\beta}^{(0,0)}g^{\mu\nu} \right) \partial_\mu \S^\alpha \partial_\nu\S^\beta   \,, \nonumber
\end{align}
which however also contain higher-order interactions.
The last line above contains the kinetic terms for the non-adiabatic sector, with $d^{(0,0)}$ and $\bar{d}^{(0,0)}$ two dimensionless symmetric real bilinear forms.
In order to avoid gradient instabilities, respectively justify us neglecting higher-order spatial derivatives, we additionally enforce $d^{(0,0)}$ to be positive, respectively definite.
Therefore, we can simultaneously reduce $d^{(0,0)}$ to the trivial identity bilinear form and co-diagonalise $\bar{d}^{(0,0)}$.
Without loss of generality, we thus consider
\begin{equation}
    d_{\alpha\beta}^{(0,0)} \rightarrow \delta_{\alpha\beta} \,, \quad\bar{d}_{\alpha\beta}^{(0,0)} \rightarrow \left(\frac{1}{c_\alpha^2}-1\right)\delta_{\alpha\beta} \quad \text{(no sum)} \,.
\end{equation}
If $\bar{d}^{(0,0)}$ is also enforced to be positive, then, $\forall \alpha\,,\, 0 \leqslant  c_\alpha^2 \leqslant 1$.
Reintroducing $\pi$ and using the transformation laws under Stückelberg, we find the following quadratic operators:
\begin{align}
\label{eq: intermeditate L2nad}
 &\sum_{\alpha}\frac{1}{2c_\alpha^2} \left[\left(\dot{\S}^\alpha\right)^2 - c_\alpha^2  \frac{\left(\partial_i \S^\alpha\right)^2}{a^2} \right]+\frac{b_{\alpha\beta}^{(0,2)}}{2}\S^\alpha\S^\beta  \\ &\quad  - c_{\alpha\beta}^{(0,1)}\dot{\S}^\alpha\S^\beta -2  b_\alpha^{(1,1)}\dot{\pi} \S^\alpha + 2 c_\alpha^{(1,0)} \dot{\pi} \dot{\S}^\alpha + \ldots
    \nonumber \,.
\end{align}
Dots denote the remaining gravitational interactions of the non-adiabatic sector with metric fluctuations in the flat gauge and can be written explicitly by the formal replacement $- 2 \dot{\pi} \rightarrow \delta g^{00}_\mathrm{flat}$ in the expression above.
Although we put those terms aside for now, we come back to them in the next section, see Eq.~\eqref{eq: grav couplings nad}.

An additional degeneracy can be broken by decomposing $c_{\alpha\beta}^{(0,1)}=c_{(\alpha\beta)}^{(0,1)} + c_{[\alpha\beta]}^{(0,1)} $ where parentheses (brackets) denote the (anti-)symmetric part of the tensor, and using integration by parts for the symmetric part,
\begin{align}
     -c_{\alpha\beta}^{(0,1)}\dot{\S}^\alpha\S^\beta =  &- c_{[\alpha\beta]}^{(0,1)}\dot{\S}^\alpha\S^\beta  - \frac{\dd}{a^3 \dd t}\left(\frac{a^3}{2}c_{(\alpha\beta)}^{(0,1)}\S^\alpha\S^\beta\right) \nonumber \\ &+\frac{\dd}{a^3 \dd t}\left(\frac{a^3}{2}c_{(\alpha\beta)}^{(0,1)}\right)\S^\alpha\S^\beta \,.
\end{align}
Dropping the total time derivative that cannot affect the dynamics nor the correlation functions of the theory (see~\cite{Braglia:2024zsl} for a detailed discussion on total time derivative terms in the in-in perturbation theory), we consistently define a symmetric mass matrix $m^2_{\S,\alpha\beta}$ of mass dimension two,
\begin{equation}
     \left[\frac{b_{\alpha\beta}^{(0,2)}}{2}+\frac{\dd}{a^3 \dd t}\left(\frac{a^3}{2}c_{(\alpha\beta)}^{(0,1)}\right)\right]\S^\alpha\S^\beta \equiv -\frac{1}{2}m^2_{\S,\alpha\beta} \S^\alpha S^\beta \,. \nonumber
\end{equation}
Further redefining mixings of mass dimension one as $b_\alpha^{(1,1)} \equiv f_\pi^2 \omega_\alpha $, $c_\alpha^{(1,0)} \equiv - (f_\pi^2/H)\bar{\omega}_\alpha $ and $c_{[\alpha\beta]}^{(0,1)}\equiv -\Omega_{\S,\alpha\beta}$ with $\Omega_{\S,\alpha\beta}$ anti-symmetric, the quadratic Lagrangian now reads
\begin{align}
\label{eq: final L2nad}
    \mathcal{L}^{(2)}_\mathrm{nad}&=\sum_{\alpha}\frac{1}{2c_\alpha^2} \left[\left(\dot{\S}^\alpha\right)^2 - c_\alpha^2  \frac{\left(\partial_i \S^\alpha\right)^2}{a^2} \right]-\frac{m^2_{\S,\alpha\beta}}{2}\S^\alpha\S^\beta \nonumber \\ &\quad  + \Omega_{\S,\alpha\beta}\dot{\S}^\alpha\S^\beta - 2  f_\pi^2 \omega_\alpha \dot{\pi} \S^\alpha - 2 \frac{f_\pi^2}{H}\bar{\omega}_\alpha \dot{\pi} \dot{\S}^\alpha 
     \,.
\end{align}

\vskip 4pt
\paragraph*{\bf Decoupling limit and quadratic action.} The total Lagrangian density is given by the sum of $\mathcal{L}_\mathrm{ad}$ in~\eqref{eq: Lad} and $\mathcal{L}_\mathrm{nad}$ in~\eqref{eq: Lnad}.
After reintroducing $\pi$, a drastic simplification occurs in the adiabatic sector by taking the so-called decoupling limit.
In this limit, defined as $\dot{H} \rightarrow 0\,,\, \Mp\rightarrow \infty$ while holding $\dot{H} \Mp^2$ fixed, the gravitational couplings of $\pi$ become negligible~\cite{Cheung:2007st}.
We here prove this statement holds true for the multi-field EFT of fluctuations under study\footnote{In Ref.~\cite{Senatore:2010wk}, it was argued that gravitational interactions become negligible in the decoupling limit even in the multi-field context, but we recall that the EFT constructed in the latter reference is less generic than ours as it relies on the non-adiabatic sector verifying a shift symmetry. Moreover, no concrete investigation of the constraints in the ADM formalism was provided (and neither in Ref.~\cite{Noumi:2012vr}), which is enough to justify our study.}.
Technically, this can be seen by using the ADM formalism (we recall that $R^{(3)}=0$ from our choice of the flat gauge), in which
\begin{align}
    \dd s^2 &= - N^2 \dd t^2 + a^2 \delta_{ij}(\dd x^i + N^i\dd t )(\dd x^j+ N^j\dd t)\,,  \\
    R&= K^{ij} K_{ij} - K^2\,,\quad K_{ij}= \frac{1}{N}\left(a^2 H \delta_{ij} - 2
    \partial_{(i}N_{j)} \right) \,, \nonumber 
\end{align}
and $\sqrt{-g} =N a^3$.
The lapse can be expanded as $N=1+\delta N$ and the shift as $N^i=\delta^{ij} \partial_i \theta /a^2 + \hat{N}^i$ with $\partial_i \hat{N}^i=0$.
The non-trivial multi-field effects concerning the decoupling limit are encoded in the gravitational mixings, 
\begin{equation}
\label{eq: grav couplings nad}
    b_\alpha^{(1,1)} \delta g^{00}_\mathrm{flat}  \S^\alpha - c_\alpha^{(1,0)}  \delta g^{00}_\mathrm{flat} \dot{\S}^\alpha \,,
\end{equation}
left as dots in~\eqref{eq: intermeditate L2nad}.
Using that $\delta g^{00}_\mathrm{flat}= 2 \delta N$, we see that only the energy constraint is affected by the presence of the non-adiabatic sector at linear order.
The momentum constraint (found from varying the action with respect to the shift) therefore reads exactly as in the purely adiabatic case,
\begin{equation}
\label{eq: lapse}
    \delta N = \epsilon H \pi + \mathcal{O}(\epsilon^2) \,,
\end{equation}
where we omitted terms of order two and more in the slow-roll expansion.
On the contrary, the energy constraint (found from varying the action with respect to the lapse) now gives
\begin{equation}
\label{eq: shift}
    \frac{\partial^2 \theta}{a^2} = - \frac{\epsilon H \dot{\pi}}{c_s^2} + \frac{b_{\alpha}^{(1,1)}}{2 \Mp^2 H}  \S^\alpha -\frac{c_{\alpha}^{(1,0)}}{2 \Mp^2 H}  \dot{\S}^\alpha + \mathcal{O}(\epsilon^2) \,,
\end{equation}
where we have identified the usual speed of sound for adiabatic fluctuations as $c_s^{-2}  = 1 + 2 M_2^4/(\Mp^2 \dot{H})$ from the adiabatic building blocks in Eq.~\eqref{eq: Lad}.
But the scalar part of the shift appears in the quadratic Lagrangian only in the combination $- 2 \Mp^2 (H \delta N + \dot{H} \pi) \partial^2 \theta/a^2$, which vanishes upon inserting the solution for the lapse, as usual.
This finishes the proof that at quadratic order in fluctuations, the only new effects of the non-adiabatic sector as far as gravitational couplings are concerned, are encoded as small corrections to the mixing interactions from inserting the solution~\eqref{eq: lapse} for the lapse in~\eqref{eq: grav couplings nad}, giving:
\begin{equation}
    -2 f_\pi^2  \omega_\alpha \left( \dot{\pi} - \epsilon H \pi \right) \S^\alpha - 2 \frac{f_\pi^2}{H} \bar{\omega}_\alpha \left( \dot{\pi} - \epsilon H \pi \right)\dot{\S}^\alpha 
     \,.
\end{equation}
It is therefore justified, in a consistent slow-roll expansion, to take the decoupling limit in the EFT of multi-field fluctuations that we have built (see Appendix B for a different viewpoint whenever the extrinsic curvature perturbation is included).
Interestingly, the slow-roll suppressed corrections from gravitational interactions are precisely the ones required to identify derivatives of the curvature perturbation from the ones of $\pi$.
Indeed, at linear order the relation between $\pi$ in the flat gauge and $\zeta$ simply reads $\zeta = - H \pi$: the quadratic mixings including gravitational interactions are exactly $(2f_\pi^2/H)( \omega_\alpha \dot{\zeta} \S^\alpha  + \bar{\omega}_\alpha\dot{\zeta} \dot{\S}^\alpha/H) $.
Since the proof is so simple, we believe there is no obstruction for extending it to any order in the fields' fluctuations as in the adiabatic case~\cite{Green:2024hbw}, but this is left for future work.

In the decoupling limit and in the flat gauge, consistently neglecting slow-roll suppressed interactions, the final quadratic Lagrangian for our multi-field EFT reads:
\begin{align}
\label{eq: final L2}
\mathcal{L}^{(2)}_\mathrm{EFT}& = \frac{f_\pi^4}{2 c_s^3}\left[ \dot{\pi}^2 - c_s^2 \frac{(\partial_i \pi)^2}{a^2}\right] \\
& \quad + \sum_{\alpha}\frac{1}{2c_\alpha^2} \left[\left(\dot{\S}^\alpha\right)^2 - c_\alpha^2  \frac{\left(\partial_i \S^\alpha\right)^2}{a^2} \right]-\frac{m^2_{\S,\alpha\beta}}{2}\S^\alpha\S^\beta \nonumber \\ &\quad  + \Omega_{\S,\alpha\beta}\dot{\S}^\alpha\S^\beta - 2 f_\pi^2  \omega_\alpha \dot{\pi} \S^\alpha - 2 \frac{f_\pi^2}{H}\bar{\omega}_\alpha \dot{\pi} \dot{\S}^\alpha \nonumber
     \,.
\end{align}
Eq.~\eqref{eq: final L2} (see also Eq.~\eqref{eq: delta L2 EFT} in Appendix B) is the main result of this work, and it is already in a final form that can be readily compared to the quadratic Lagrangian for fluctuations in NLSM, see Eq.~\eqref{eq: L2 model}.
Although definitely similar, we now highlight their differences.
The first line in Eq.~\eqref{eq: final L2} concerns the adiabatic sector only and contains a speed of sound for $\pi$.
The second line shows that non-adiabatic fluctuations may all have speeds of sound  deviating from unity too.
The third line describes quadratic kinetic mixings, in particular $\dot{\pi}$ is coupled via $\omega_\alpha$ to all fluctuations $\S^\alpha$ (see Appendix C for a related discussion on flavour and mass bases in the multi-field EFT) and, via $\bar{\omega}_\alpha$, also to their time derivatives.
The quadratic mixing operator $\propto \bar{\omega}_\alpha$ is also absent in two-field DBI inflation and contrary, for example, to the speeds of sound which indeed arise in this model~\cite{Langlois:2008qf}.
We are not aware of any concrete model of inflation that features it, and this is both the strength and the weakness of the EFT approach: we can systematically propose operators allowed by symmetries and go beyond the lamppost of known models, but the resulting EFT may not have a motivated UV realisation.

\vskip 4pt
\paragraph*{\bf Non-linearly realised symmetries and cubic interactions.}

Many terms can be written at the cubic order in fields' fluctuations from the non-adiabatic blocks~\eqref{eq: Lnad}.
Instead of listing all of them, we here focus on the subset of those that are fixed by non-linearly realised spacetime symmetries.
The existence of the latter, most famously relating the speed of sound $c_s^2$ to the strength of the cubic interactions $\dot{\pi} (\partial \pi)^2$ in the adiabatic sector~\cite{Cheung:2007st}, is well known.
Less studied are their consequences in the multi-field context (see however~\cite{Senatore:2010wk} for examples where the matter content verifies an additional shift symmetry and the recent works~\cite{Werth:2023pfl,Pinol:2023oux} taking advantage of these symmetries in specific two-field realisations).
Here we list all cubic multi-field operators whose strengths are fixed by operators already present in the quadratic Lagrangian.
\begin{itemize}
    \item The operator $b^{(1,1)}_\alpha \delta g^{00} \S^\alpha$ responsible for the quadratic mixing $\propto \omega_\alpha$ brings the following cubic order contribution which cannot be reproduced by other operators:
    \begin{equation}
         f_\pi^2\omega_\alpha \frac{(\partial_i \pi)^2}{a^2}\S^\alpha \,.
    \end{equation}
    Note that we have not highlighted the other contribution $-f_\pi^2\omega_\alpha \dot{\pi}^2\S^\alpha $ as it can also be generated from an independent EFT building block, namely the one $\propto b_\alpha^{(2,1)}$ in~\eqref{eq: Lnad}.
    \item The operator $c_{\alpha\beta}^{(0,1)} g^{0\mu} \partial_\mu \S^\alpha\S^\beta$ whose anti-symmetric component gives the non-adiabatic sector mixing $\propto \Omega_{\S,\alpha\beta}$ also fully fixes
        \begin{equation}
         -\Omega_{\S,\alpha\beta} \frac{\partial_i \pi \partial_i \S^\alpha}{a^2 } S^\beta\,.
    \end{equation}
    \item  The operator $c_\alpha^{(1,0)} \delta g^{00} g^{0\mu} \partial_\mu \S^\alpha $ giving the new mixing terms $\propto \bar{\omega}_\alpha$ also contributes uniquely as
    \begin{equation}
    \label{eq: NLRS bar omega}
        \frac{f_\pi^2}{H}\bar{\omega}_\alpha\frac{(\partial_i \pi)^2}{a^2}\dot{\S}^\alpha - 2\frac{f_\pi^2}{H} \bar{\omega}_\alpha\dot{\pi}\frac{\partial_i \pi \partial_i \S^\alpha}{a^2}\,.
    \end{equation}
    \item  The operator with $d_{\alpha\beta}^{(0,0)}$ is a four-dimensional scalar and transforms covariantly under time diffeomorphisms, while the one with $\bar{d}_{\alpha\beta}^{(0,0)}$ giving the speeds of sound for the non-adiabatic fluctuations also results in an unique cubic operator
    \begin{equation}
    \label{eq: before time dependence discussion}
        -\sum_{\alpha}\left(\frac{1}{c_\alpha^2}-1\right) \frac{\partial_i \pi \partial_i \S^\alpha}{a^2} \dot{\S}^\alpha \,.
    \end{equation}
\end{itemize}
In addition to these spatial derivative interactions, all couplings being a priori time-dependent, they should be consistently expanded at the time $t+\pi$ after the transformation, giving for any quadratic operator $\mathcal{O}_2$, $c(t)\mathcal{O}_2 \rightarrow \dot{c}(t)\pi \mathcal{O}_2$ at cubic order.
For example, the quadratic mixing $\propto \omega_\alpha$ is found again at cubic order as $- 2 f_\pi^2 \dot{\omega}_\alpha \dot{\pi} \pi \S^\alpha$, and here we generalised this to all time-dependent quadratic couplings of the EFT in Eq.~\eqref{eq: final L2}.
Cubic operators resulting from this time shift are also uniquely fixed, as no non-derivative operator $\pi$ can be introduced from other contributions in the unitary gauge (except slow-roll suppressed gravitational couplings via the lapse $\delta N$).
Symmetries dictating cubic interactions as proportional to time derivatives of quadratic mixings are crucial, as they can lead to correlated features in the primordial power spectrum and bispectrum~\cite{Chen:2011zf,Werth:2023pfl,Pinol:2023oux}.
When applicable to NLSM, i.e. with $\omega_{\alpha} =\delta_{\alpha 1} \omega_1$ and $\bar{\omega}_\alpha = 0 = 1/c_\alpha^2-1$, we check in Appendix A that all those cubic interactions fixed by symmetries are indeed explicitly present with the exact same size, which is an important and non-trivial consistency check between those two independent descriptions of multi-field interactions.

\vskip 4pt
\paragraph*{\bf Discussion.}

By considering any number of non-adiabatic fluctuations, and without assuming additional symmetries in the matter sector, we constructed the general effective field theory of multi-field inflationary fluctuations.
We proposed adiabatic and non-adiabatic building blocks verifying the unbroken FLRW symmetries in the unitary gauge, removed degenerate operators after reintroducing $\pi$ with the Stückelberg trick, and identified the full quadratic Lagrangian pertinent for deriving model-independent predictions in multi-field inflation.
This was done consistently in a derivative and slow-roll expansion and, using the ADM formalism to solve for the constraints, we proved explicitly that in the decoupling limit all gravitational interactions are small and can safely be neglected.
We showed how non-linearly realised spacetime symmetries can be used to fix the sizes of several multi-field cubic couplings, akin to the purely adiabatic case.
We checked explicitly that all these couplings protected by symmetries were indeed found in concrete realisations known as non-linear sigma models, showing consistency and synergy between these two complementary approaches.

We expect our EFT construction to stimulate new phenomenological searches for signals of multi-field inflation in a model-independent way.
Directions for future work include the cosmological collider signal mediated by the new mixing $\propto \bar{\omega}_\alpha$, effects due to the speeds of sound $c_\alpha^2$, and couplings with tensor modes.
On the theoretical side, our EFT was derived in consistent expansions in derivatives and field's fluctuations, and allows for straight generalisations to higher orders in derivatives and to quartic interactions also fixed by symmetries.
Moreover, we have considered here a quasi de Sitter spacetime, while the effects of even tiny departures from slow variation may leave interesting imprints in a multi-field context, and these scenarios too can be explored with the general tool provided in this work.
Finally, it would be interesting to apply our EFT to other scalar fluctuations in cosmology, beyond the inflationary context.
    
\vskip 4pt
\begin{acknowledgments}
I would like to thank S. Garcia-Saenz for regular insightful discussions related to the EFT treatment of the adiabatic sector, as well as M. Braglia, T. Noumi and D. Werth for useful comments on a draft of this manuscript.
I also acknowledge funding support from the Initiative Physique des Infinis (IPI), a research training program of the Idex SUPER at Sorbonne Université.

\end{acknowledgments}

\bibliography{bib}

\subfile{supplemental}

\end{document}

%% file: supplemental.tex
\section*{Supplemental Material}

\vskip 4pt
\paragraph*{\bf Appendix A. Cubic interactions in non-linear sigma models.}

The cubic Lagrangian density of fluctuations in non-linear sigma models is composed of many interactions, amongst which notable ones read~\cite{Pinol:2020kvw}:
\begin{align}
    \mathcal{L}^{(3)}_\mathrm{NLSM}\supset \, &  \, \frac{f_\pi^2}{H^2}  \delta_{\alpha 1} \left\{\omega_1 \left[\frac{(\partial_i \zeta)^2}{a^2}- \dot{\zeta}^2\right] -  2 \dot{\omega}_1 \dot{\zeta} \zeta \right\} \F^\alpha  \nonumber \\
    & + \frac{1}{H} \left[ \Omega_{\alpha\beta} \frac{\partial_i\zeta\partial_i\F^\alpha}{a^2} \F^\beta -  \dot{\Omega}_{\alpha\beta} \zeta\dot{\F}^\alpha \F^\beta \right] \nonumber \\
    & +\frac{\dot{m}^2_{\alpha\beta}}{2H} \zeta \F^\alpha \F^\beta \\
    & - \frac{1}{6}V_{;\alpha\beta\gamma}\F^\alpha\F^\beta\F^\gamma + \frac{2  f_\pi^2}{3} R_{\alpha\beta\gamma\sigma}\dot{\F}^\alpha\F^\beta\F^\gamma \,. \nonumber 
\end{align}
The first line contains the interactions fixed by the non-linearly realised symmetry in the EFT approach for a mixing of the form $\omega_\alpha = \delta_{\alpha 1} \omega_1$, but note that an additional fine-tuning is present in NLSM: the parameter $b^{(2,1)}$ seems to be vanishing, as otherwise it would additionally break the symmetry fixing the size of the $\dot{\zeta}^2 \F^1$ interaction.
The second line is not explicitly present in the final Lagrangian in~\cite{Pinol:2020kvw} as it has been expressed in terms of other cubic interactions, however it can be found at an intermediate stage in Eq.~(3.7) of the latter reference, and contains exactly the cubic interactions fixed by the presence of the kinetic mixing operator in the purely non-adiabatic sector.
The third line contains the interaction expected from the EFT viewpoint given the presence of the mass matrix evaluated at $t+\pi$ after Stückelberg.
As already known from the study of the quadratic Lagrangian, there is no cubic interaction related to the presence of speeds of sound, either from $\zeta$ or the entropic fluctuations.
Actually, in NLSM, there is no cubic interaction at all that involve a total of three derivatives as in Eqs.~\eqref{eq: NLRS bar omega}--\eqref{eq: before time dependence discussion}, or from operators $\propto c^{(2,0)}$ and $\bar{d}^{(1,0)}$, and contrary to, e.g., the two-field DBI scenario~\cite{Langlois:2008wt}.
We therefore suspect that all operators with $c^{(n,0)}$ and $\bar{d}^{(n,k)}$ are absent in NLSM at every order of perturbation theory.
In order to contrast with these terms fixed by symmetries, we show in the fourth line two cubic interactions that we do not expect to be fixed from the EFT view point.
These coefficients are projections of the covariant third derivatives of the potential and of the Riemann tensor of the internal field space, which are both absent from the quadratic Lagrangian. 
Although these terms cannot, indeed, be predicted from the knowledge of the quadratic Lagrangian in the EFT, interactions of these forms are expected from the presence of building blocks $b^{(0,3)}$, respectively $c^{(0,2)}$, showing once more consistency and synergy between the two different approaches.

\vskip 4pt
\paragraph*{\bf Appendix B. Including three-dimensional curvatures.} We explore the effects of including in the EFT operators made off the 3d extrinsic curvature perturbation $\delta K_{ij}$ and the intrinsic one $R^{(3)}_{ij}$. 
We first review known features in the adiabatic sector and then dig into new ones in the non-adiabatic sector.
We will need the rules of transformation of these objects under the Stückelberg trick restoring diffeomorphism invariance,
\begin{align}
        \delta K_{ij} & \rightarrow \delta K_{ij}^\mathrm{flat} + a^2 \epsilon H^2 \pi \delta_{ij} - \partial_i \partial_j \pi + \ldots  \,, \\
    R^{(3)}_{ij} & \rightarrow H \left(\partial_{i}\partial_j \pi + \delta_{ij} \partial^2 \pi \right) +\ldots \,, \nonumber
\end{align}
where we recall that we chose to define $\pi$ in the flat gauge, so $R^{(3),\mathrm{flat}}_{ij}$ vanishes identically.
The dots represent terms of higher order in field's fluctuations and, together with slow-roll suppressed terms, we will neglect them in this discussion.

\subparagraph{In the adiabatic sector.}
We consider the addition of the following towers of EFT building blocks in the adiabatic sector and in the unitary gauge,
\begin{align}
     \Delta \mathcal{L}_\mathrm{ad} = & \, \sum_{n=1}^\infty\frac{1}{2n!}\left(\delta g^{00}\right)^n \left[\bar{M}_n^3  \delta K + m_n^2 R^{(3)}  \right] \\
     & \, + \sum_{n=0}^\infty\frac{1}{n!} \tilde{m}_n^2 \left(\delta g^{00}\right)^n \left(\delta K^{ij} \delta K_{ij} - \delta K^2  \right) \nonumber + \ldots
\end{align}
Adding these operators to Eq.~\eqref{eq: Lad} leads to the most general Lagrangian for the adiabatic degree of freedom with linear equations of motion of order strictly two\footnote{
The combination $\delta K^{ij}R^{(3)}_{ij}-\delta K R^{(3)}$ also brings terms with a single derivative per fluctuation only, however it is redundant with other operators quadratic in the 3d curvatures as shown in~\cite{Gleyzes:2013ooa}, and can therefore be dismissed without loss of generality.
}, both in time and space, therefore encompassing the linearised version of general single-field theories named after Horndeski~\cite{Horndeski:1974wa} and rediscovered as ``generalised galileons''~\cite{Deffayet:2009mn,Deffayet:2011gz}, but however surpassing them whenever $m_1^2 \neq \tilde{m}_0^2 $~\cite{Gleyzes:2013ooa}.
The dots in the expression above represent operators that eventually lead to terms with more than one derivative per field's fluctuations and can be consistently neglected in the derivative expansion, with the exception of degenerate kinetic terms like in the ghost condensate scenario~\cite{Arkani-Hamed:2003pdi}.
Actually, the combination $\tilde{m}_0^2\left(\delta K^{ij} \delta K_{ij} - \delta K^2  \right)$ can be removed by the combination of a disformal and a conformal transformations of the metric~\cite{Gleyzes:2014rba} after which gravitational waves have unit sound speed of propagation~\cite{Creminelli:2014wna}.
In the following we directly work in this frame pertinent for comparison of theoretical predictions with observations, and we therefore assume that this operator has been removed.
Being mainly interested in the quadratic Lagrangian density, we therefore focus on the following two operators giving, after Stückelberg and at quadratic order,
\begin{align}
\label{eq: 3d curvatures after Stuckelberg}
    \frac{\bar{M}_1^3}{2} \delta g^{00} \delta K &\rightarrow \frac{\bar{M}_1^3}{2} \left( \delta g^{00}_\mathrm{flat} - 2 \dot{\pi} \right) \left(\delta K_\mathrm{flat} - \frac{\partial^2 \pi}{a^2} \right)  \,, \nonumber  \\
    \frac{m_1^2}{2}   \delta g^{00} R^{(3)} &\rightarrow \frac{m_1^2}{2}   \left( \delta g^{00}_\mathrm{flat} - 2 \dot{\pi} \right) 4 H \frac{\partial^2 \pi}{a^2}\,,
\end{align}
where we kept explicitly all possible gravitational couplings.
Repeating arguments of the main text and since $\delta g^{00}_\mathrm{flat} = 2 \delta N$ in the ADM formalism, we see that $m_1^2$ does not enter the solution for the lapse.
On the contrary, since $\delta K_\mathrm{flat} = -3 H \delta N  -  \partial^2 \theta/a^2$, $\bar{M}_1$ does change the lapse at linear order, explicitly,
\begin{align}
    \delta N &= \frac{1}{1+\alpha} \epsilon H \pi + \frac{\alpha}{1+\alpha}\dot{\pi} + \mathcal{O}(\epsilon^2) \,, \,\, \text{with} \\ \alpha 
    &= \frac{\bar{M}_1^3}{2H \Mp^2} \,. \nonumber
\end{align}
Plugging this solution in the total quadratic Lagrangian density, we confirm once more that the shift contributions cancel, and we get new contributions of the forms $\pi^2, \pi \dot{\pi}, \dot{\pi}^2, \pi \partial^2 \pi, \dot{\pi} \partial^2 \pi$ both directly from the operators in Eq.~\eqref{eq: 3d curvatures after Stuckelberg} and via all gravitational couplings including $\delta N$ above, both kinds of contributions being of similar order of magnitude.
By time and space integration by parts they can all be recast in three operators only: $\pi^2$ whose final coefficient cancels at leading order in slow roll as expected, $\dot{\pi}^2$ with coefficient $c_1$ and $(\partial \pi)^2/a^2$ with coefficient $c_2$, thus defining a speed of sound $c_s^2 = c_2 / c_1$, see e.g.~\cite{Gleyzes:2013ooa} for the exact expressions.

We now comment on the possible size of the corrections induced by these two new operators linear in the 3d curvatures in the unitary gauge, depending on assumptions about the UV realisation of the EFT.
If the non-universal operators are all fixed by a single energy scale $\Lambda$ defining the cutoff of the EFT, e.g. if the parent theory is of the form $\Lambda^4 \left[P(X/\Lambda^4, \phi/\Lambda) + G(\square \phi/\Lambda^3, \phi/\Lambda) + \ldots \right]$, then all Wilson coefficients parametrically read $M_2 \sim \bar{M}_1 \sim m_1 \sim \Lambda \sim \sqrt{H \Mp}$ where the last approximate relation comes from the first Friedmann equation.
In this case, the correction from $\bar{M}_1$ in $\delta N$ is suppressed as $\alpha \sim \sqrt{H/\Mp}$, and taking the decoupling limit is justified.
Actually, in this regime, it is easy to check that all contributions from the operators linear in the 3d curvatures are suppressed by positive powers of $\sqrt{H/\Mp}$ compared to $M_2^4 (\delta g^{00})^2$, therefore justifying us neglecting them in the main body of this manuscript for this class of UV realisations.
However, it may well happen that there exist several high-energy scales, as is the case for example in weakly broken galileons~\cite{Pirtskhalava:2015nla, Pirtskhalava:2015zwa}, where $\Lambda_2$ appears in first-order derivative terms like $X/\Lambda_2^4$, and $\Lambda_3$ appears in second-order ones like $\square \phi /\Lambda_3^3$ with the relation $\Lambda_2^4 = \Mp \Lambda_3^3$.
If, e.g., $M_2 \sim \Lambda_2$ and $\bar{M}_1 \sim \Lambda_3$ then $\alpha \sim M_2^4/(H^2\Mp^2) \sim 1$ and one has to consider all contributions to the speeds of sound including gravitational couplings, as they may all bring order-one corrections.
To put it in a nutshell, we have understood that, in the adiabatic sector, terms including the 3d curvatures are either negligible as expected from a naive derivative expansion in terms of the metric fluctuations, or may bring sizeable corrections depending on assumptions about the UV realisation of the EFT and, in the latter case, taking the decoupling limit is not justified and constraints must be solved and taken into account.

\subparagraph{In the non-adiabatic sector.}
Based on the understanding in the adiabatic sector, we consider the addition of towers linear in the 3d curvatures in the non-adiabatic sector and in the unitary gauge, namely
\begin{align}
     \Delta \mathcal{L}_\mathrm{nad} = & \, \sum_{n,k}^\infty\frac{1}{n!k!}\left(\delta g^{00}\right)^n \S^{\alpha_1} \ldots \S^{\alpha_k}   \\ 
     & \times \Bigl[\bar{b}^{(n,k)}_{\alpha_1 \dots \alpha_k} \delta K + \tilde{b}^{(n,k)}_{\alpha_1 \dots \alpha_k}  R^{(3)} \nonumber \\
     & + \bar{c}^{(n,k)}_{\alpha \alpha_1 \dots \alpha_k} \delta K  g^{0\mu} \partial_\mu \S^\alpha    + \tilde{c}^{(n,k)}_{\alpha \alpha_1 \dots \alpha_k} R^{(3)} g^{0\mu} \partial_\mu \S^\alpha \nonumber \\
     &  +\ldots \,
     \Bigr] \nonumber \,,
\end{align}
where we omitted terms of higher-derivative order.
The sums are on $(n,k) \in \mathbb{N}^2$, but we fix $\bar{b}^{(0,0)}=\tilde{b}^{(0,0)}=0$ to avoid tadpoles, as well as more generally all $\bar{b}^{(n,0)}=\tilde{b}^{(n,0)}=0$ to avoid degeneracies with operators $\propto \bar{M}_n^3 \,, m_n^2$ defined in the adiabatic sector.
Being mainly interested in the quadratic Lagrangian density, we specifically focus on the following four operators giving, after Stückelberg and at quadratic order,
\begin{align}
\label{eq: 3d curvatures after Stuckelberg nad}
    \bar{b}^{(0,1)}_\alpha\delta K  \S^\alpha  &\rightarrow \bar{b}^{(0,1)}_\alpha \left(\delta K_\mathrm{flat} - \frac{\partial^2 \pi}{a^2} \right)  \S^\alpha \,,  \\
     \tilde{b}^{(0,1)}_\alpha R^{(3)}  \S^\alpha  &\rightarrow \tilde{b}^{(0,1)}  4 H \frac{\partial^2 \pi}{a^2} S^\alpha \nonumber \,,  \\
     \bar{c}_\alpha^{(0,0)} \delta K  g^{0\mu} \partial_\mu \S^\alpha  &\rightarrow  - \bar{c}_\alpha^{(0,0)} \left(\delta K_\mathrm{flat} - \frac{\partial^2 \pi}{a^2} \right) \dot{\S}^\alpha \nonumber \,, \\
     \tilde{c}_\alpha^{(0,0)} R^{(3)}  g^{0\mu} \partial_\mu \S^\alpha  &\rightarrow  - \tilde{c}_\alpha^{(0,0)}  4 H \frac{\partial^2 \pi}{a^2} \dot{\S}^\alpha \nonumber \,.
\end{align}
The operators proportional to tilde coefficients are already in a final form and do not affect the constraint equations and, contrary to the adiabatic sector, they cannot be rewritten in terms of other quadratic operators already present in the multi-field Lagrangian.
Therefore, they bring genuinely new interactions to the ones shown in the main body of this manuscript.
This is important as, e.g., the operator $\partial^2 \pi \S^\alpha$ was encountered in models of inflation where $\S^1$ represents a scalar degree of freedom of an additional spin-two fluctuation~\cite{Bordin:2018pca}, and predicted in the two-field EFT~\cite{Noumi:2012vr}, so they must appear in our EFT too (note that they are absent from NLSM).
The operators proportional to barred coefficients need more care as they will affect the solution for the lapse via the shift appearing in $\delta K_\mathrm{flat}$, and their gravitational couplings cannot be neglected a priori.
At linear order, the lapse now reads (we do not consider extrinsic curvature terms in the adiabatic sector here, i.e. we tune $\bar{M}_1=0$ for simplicity)
\begin{align}
\delta N &= \epsilon H \pi + \beta_\alpha \S^\alpha  + \gamma_\alpha \dot{\S}^\alpha + \mathcal{O}(\epsilon^2) \,, \,\, \text{with} \\ 
\beta_\alpha &= - \frac{ \bar{b}^{(0,1)}_\alpha}{2\Mp^2 H} \quad \text{and} \quad \gamma_\alpha = \frac{ \bar{c}^{(0,0)}_\alpha}{2\Mp^2 H} \,. \nonumber
\end{align}
The solution for the shift is also modified, but its final contributions cancel each other, as usual.
One can check that the non-vanishing contributions from the operators proportional to barred coefficients are of the forms $\pi \S^\alpha, \pi \dot{\S}^\alpha, \partial^2 \pi \S^\alpha,  \partial^2 \pi \dot{\S}^\alpha, \S^\alpha \S^\beta, \S^\alpha \dot{\S}^\beta, \dot{S}^\alpha \dot{S}^\beta$.
Although a priori all important, they are degenerate with other quadratic terms of the multi-field EFT.

To put it in a nutshell, in this appendix we have shown that, whenever 3d spatial curvatures are included in the multi-field EFT, two additional quadratic terms arise,
\begin{equation}
\label{eq: delta L2 EFT}
    \Delta \mathcal{L}^{(2)}_\mathrm{EFT} = - \rho_\alpha \frac{f_\pi^2}{H} \frac{\partial^2 \pi}{a^2 } \S^\alpha  - \bar{\rho}_\alpha \frac{f_\pi^2}{H^2} \frac{\partial^2 \pi}{a^2 } \dot{\S}^\alpha\,,
\end{equation}
with $\rho_\alpha$ and $\bar{\rho}_\alpha$ couplings of mass dimension one. 
The second term has more than one derivative per field and should consistently be neglected at low energies.
But the first one, that can be rewritten as proportional to $\partial_i \pi \partial_i \S^\alpha$, comes at the same derivative order as the interaction $\dot{\pi}\dot{\S}^\alpha$\footnote{
Note that both operators, $\dot{\pi} \dot{\S}^\alpha$ and $\partial_i\pi\partial_i \S^\alpha$, behave as purely kinetic mixings.
In particular, when their wave-numbers are still deep in the sub-Hubble regime, they do not become negligible in contrast with the adiabatic-entropic mixing $\dot{\pi} \S^\alpha$ and the mass and kinetic mixings in the purely entropic sector.
Therefore, they affect the dispersion relation and can potentially lead to instabilities.
In turn, requiring stability of the theory puts upper bounds on the couplings $\bar{\omega}_\alpha$ (see~\cite{Baumann:2011nk}) and $\rho_\alpha$ (see~\cite{Bordin:2018pca}).
} and is both found in inflationary scenarios with spinning fields~\cite{Bordin:2018pca} and expected from the two-field EFT construction in~\cite{Noumi:2012vr}.
Additionally, this new operator also breaks some of the standard predictions for cubic interactions from non-linearly realised symmetries shown in the main body, a feature already noticed in the adiabatic sector~\cite{Pirtskhalava:2015zwa}.

\vskip 4pt
\paragraph*{\bf Appendix C. Flavour and mass bases.}

In this appendix, we show how unequal speeds of sound $c_\alpha \neq c_\beta$ forbid in general to define the flavour and mass bases found in non-linear sigma models~\cite{Pinol:2021aun}.
For simplicity, we fine-tune $\bar{\omega}_\alpha = 0$ in this discussion.
The flavour basis is defined as the one in which non-adiabatic fluctuations have i) diagonal kinetic terms; ii) mixings with the adiabatic sector in a particular symmetric form $-2f_\pi^2\omega_\alpha\dot{\pi} \S^\alpha \rightarrow -2f_\pi^2 \delta_{\alpha 1} \omega_1 \dot{\pi} \F^\alpha$, see Eq.~\eqref{eq: L2 model}.
The mass basis corresponds to the same first criterion i), but with the second one being ii)bis the mass matrix is diagonal with $m^2_{\S,\alpha\beta} \S^\alpha \S^\beta \rightarrow \sum_{\alpha} m_\alpha^2 (\sigma^\alpha)^2$.

For the flavour basis, a natural possibility consists in rotating the orthonormal basis $\S^\alpha$ to pick a preferred direction as specified by the mixing $\omega_\alpha$.
We define
\begin{equation}
    \F^1 \equiv \frac{\omega_\alpha \S^\alpha}{\sqrt{\sum_\beta \omega_\beta^2}} \,,
\end{equation}
and all $\F^{\alpha \geqslant 2}$ from the knowledge of this $\F^1$ and the remaining $\S^{\alpha \geqslant 2}$ using the Gram-Schmidt algorithm.
Because both bases are orthonormal, they are simply related by a rotation matrix $R$ such that $\F^\alpha= R^\alpha{}_\beta \S^\beta$.
In this new basis, the mixing with the adiabatic sector indeed takes the desired form for criterion ii), and the mass matrix and other mixings nicely transform under the change of basis.
The gradient terms being independent of $c_\alpha$, they simply give $-\sum_\alpha (\partial_i \F^\alpha)^2/2$.
But the time derivatives are crucially affected, in the new basis they read (overlooking terms with derivatives of the rotation matrix that can be incorporated in the mixing matrices) 
\begin{equation}
    \sum_{\gamma} \frac{1}{2 c_\gamma^2}
    R^{\gamma}{}_\alpha R^{\gamma}{}_\beta \dot{\F}^\alpha \dot{\F}^\beta \,.
\end{equation}
The resulting kinetic term is therefore not diagonal, as can be seen explicitly by specifying to a two-dimensional rotation matrix with parameter $\varphi$, which gives a canonical kinetic term with speeds of sound $c_1$, plus $(c_2^2-c_1^2)/(2c_1^2c_2^2)\times [\mathrm{cos}(\varphi)\dot{\F}^1+\mathrm{sin}(\varphi)\dot{\F}^2]^2 $, therefore contradicting criterion i) for the definition of the flavour basis.

As for the mass basis, we can indeed diagonalise the mass matrix $m^2_{\S,\alpha\beta}$ with eigenvectors $\sigma^\alpha$.
As it is real and symmetric, the matrix of change of basis $\S^\alpha \rightarrow \sigma^\alpha$ is again a rotation.
Although the mass terms have now the desired form for criterion ii)bis and the kinetic mixings again transform nicely, the time derivative kinetic terms violate condition i) again, from the same reason as above.

We conclude about the impossibility in the multi-field EFT to simultaneously co-reduce the two bilinear forms appearing as kinetic terms and, either single out a preferred direction for the mixing as $-2f_\pi^2 \delta_{\alpha 1} \omega_1 \dot{\pi} \F^\alpha$, or co-diagonalise the mass matrix, except for the sub-class of models where all non-adiabatic sound speeds are equal, i.e. if $\forall \alpha\,,\, c_\alpha^2 = c^2$.
Non-linear sigma models of inflation are precisely falling into this class, thus making the striking phenomenology of inflationary flavour oscillations unveiled in Ref.~\cite{Pinol:2021aun} a unique feature in the landscape of multi-field effective field theories.